\documentclass[conference]{IEEEtran}

\usepackage[english]{babel}
\usepackage{ifthen}
\usepackage{xspace}
\usepackage{subfigure}
\usepackage{booktabs}
\usepackage{ifthen}
\usepackage{amsfonts}
\usepackage{amssymb}
\usepackage{graphicx}
\usepackage[cmex10]{amsmath}
\usepackage{alltt}
\usepackage{xfrac}
\usepackage{fancyvrb}

\newboolean{showcomments}
\setboolean{showcomments}{false}
\ifthenelse{\boolean{showcomments}}{%
   \newcommand{\bnote}[2]{
    \fbox{\bfseries\sffamily\scriptsize #1}
    {\sffamily\small$\blacktriangleright$#2$\blacktriangleleft$}}
}{%
   \newcommand{\bnote}[2]{}
}

\newcommand{\ie}{i.e.,\xspace}
\newcommand{\eg}{e.g.,\xspace}
\newcommand{\etal}{et al.\xspace}
\newcommand{\etc}{etc.\xspace}

\newcommand{\secref}[1]{Section~\ref{#1}\xspace}

\newcommand{\Figref}[1]{Fig.~\ref{#1}\xspace}
\newcommand{\tabref}[1]{Table~\ref{#1}\xspace}

\newcommand{\equref}[1]{Equation~\eqref{#1}\xspace}

\newcommand{\myparagraph}[1]{\noindent\textbf{#1}}

\newcommand{\interfaces}{\mathcal{I}}
\newcommand{\signatures}{\mathcal{S}}

\newcommand{\interface}{i}

\newcommand{\iSize}{size(\interface)}

\newcommand{\A}{{\tt \textsc{A}}\xspace}
\newcommand{\B}{{\tt \textsc{B}}\xspace}
\newcommand{\C}{{\tt \textsc{C}}\xspace}
\newcommand{\D}{{\tt \textsc{D}}\xspace}
\newcommand{\E}{{\tt \textsc{E}}\xspace}
\newcommand{\F}{{\tt \textsc{F}}\xspace}

\newenvironment{mydescription}{%
    \begin{description} \setlength{\itemsep}{0cm}}
    {\end{description}}

\newcounter{rq}
\newcounter{q}
\newenvironment{Q}{
	\refstepcounter{q}
	\begin{mydescription}
		\item[{\large Q}\arabic{q}]
		}
	{\end{mydescription}}
\newcommand{\refq}[1]{Q\ref{#1}}

\newcounter{mydef}
\begin{document}

\title{Software Interfaces: \\ On The Impact of Interface Design Anomalies\\ 
\vspace{0.1cm}
\footnotesize Author manuscript -- published in IEEE CSIT' 2013 -- DOI: 10.1109/CSIT.2013.6588778}

\author{
	\IEEEauthorblockN{
		Hani Abdeen,  
		Osama Shata, and  
		Abdelkarim Erradi
		}
	\IEEEauthorblockA{
		Department of Computer Science Engineering, 
		Qatar University, Doha, Qatar
		}
}

\maketitle

\begin{abstract}
Interfaces are recognized as an important mechanism to define contracts governing interactions between semi-independent software modules. 
Well-designed interfaces significantly reduce software complexity and ease maintainability by fostering modularization, hiding implementation details and minimizing the impact caused by changes in the software implementation. 
However, designing good interfaces is not a trivial task. 
The presence of interface design defects often yield increased development cost, lower code quality and reduced development productivity. 
Despite their importance, currently there are only a few research efforts that investigate the quality of interface design. 
In this paper, we identify and characterize common interface design anomalies and illustrate them via examples taken from well-known open source applications. 
In order to quantify the presence of interface design anomalies and estimate their impact on the interface design quality, as well on the software quality attributes, such as maintainability, we conduct an empirical study covering 9 open source projects. 
Building on our empirical results, we develop a set of recommendations to improve interface design.     

\end{abstract}


\vspace{-0.3cm}
\section{Introduction}\label{sec:introduction}
Interfaces represent abstract service contracts governing interactions between semi-independent software modules. 
In object-oriented software systems, they are reference types used to encode shared services among classes of different types \cite{Bill98a,Warr99a}. 
They are also used to define the system \textit{Application Programming Interfaces} (API) \cite{henning2007api,Dig2005}. 
Well-designed interfaces eases reusability, maintainability and testability of the software \cite{henning2007api,Dani11a,Dig2005,Bill98a,Warr99a}. 
Interfaces should be intuitive and easy to understand, however, designing good interfaces is not a trivial task \cite{henning2007api}. 
It is rather a sensitive task with a large influence on the rest of the system \cite{Dani11a,Dig2005}.

However, as software evolves over time with the modification, addition and removal of new classes and interfaces, the software gradually drifts and looses quality \cite{Eick01a}. 
To help maintainers improve the software quality, there has been recently an important progress in the area of object oriented refactoring and automatic optimization of code quality \cite{Mens04b}. Most of the existing approaches are mainly based on source code metrics, such as the ones defined by Chidamber and Kemerer \cite{Chid94a}, and the predefined bad smells in source code suggested by Fowler and Beck \cite{Fowl99a}.   
In spite of their good performance, none of these approaches take into account the particularities of Software interfaces. 

In the literature, few recent works attempt to address the particularities of interfaces.
There are some well-known interface design principles, such as {\it ``Program to an interface, not an implementation''} so as to reduce coupling, and the Interface Segregation Principle (ISP) i.e., avoid `fat' and non-cohesive interfaces that serve different clients \cite{Mart00b}. 
Currently there are only a few research efforts that investigate the quality of interfaces design. 
Most existing work largely investigated the detection of design anomalies at the class level without focusing on the specifics of interfaces \cite{Dani11a}. 
Studying the impact of interface design anomalies on software quality has been mostly neglected. 
This paper aims to address this gap as it constitute a potential barrier towards fully realizing the benefits of using interfaces as a central element for modular design.

\myparagraph{Contributions.} 
In this paper, we investigate and characterize two interface design anomalies: 
(1) duplicate interface methods (i.e., methods that are redundantly declared in different interfaces); and 
(2) unused interface methods. 
We illustrate these design anomalies via examples taken from well-known open source applications. 
Then we empirically assessed their impact on software quality.
We conduct an empirical study covering 9 open source projects in order to quantify the presence of those interface design anomalies and estimate their impact on the cohesiveness of software interfaces, as well on the software complexity and maintainability. 
The results of our study show that our proposed interface design anomalies are common and present in real systems at different degrees. 
They demonstrate that software engineers tend to overuse interfaces for declaring unnecessary methods that may even leak the internal behavior of implementing classes. 
Furthermore, they point that method clones are widely present in interfaces. 
Such interface methods also cause the degradation of interface usage cohesion.

The remainder of the paper is organized as follows. 
\secref{sec:relatedWorks} discusses existing works related to interface design quality assessment. 
We characterize two common interface design anomalies in \secref{sec:designAnomalies}. 
Then in \secref{sec:setup} we describe our empirical experiments conducted to study the identified interface design anomalies through an exploratory study that aims at answering different research questions. 
\secref{sec:results} presents and analyses the results of the empirical study. 
We discuss the results of the conducted empirical study in \secref{sec:discussion}. 
Finally, we present treats to their validity in \secref{sec:threats} before concluding. 

\section{Related Works}\label{sec:relatedWorks} 

Previous works related to software design anomalies mainly revolve around code quality, code smells, and refactoring support for object-oriented applications. Few recent works attempt to address the particularities of interface design.

\subsection{Design Defects Detection and Correction}\label{sec:relatedWorksRefactoring}
\vspace{-0.15cm}
In recent years there has been considerable interest in automatic detection and correction of design defects in object oriented software \cite{Fowl99a,Matthew2005,Trif05b,Liu12b,Hill12b}.
Mens and Tourw\'e \cite{Mens04b} survey shows that existing approaches are mainly based on code metrics and predefined bad smells in source code \cite{Chid94a,Hend96a,Basi96b,Fowl99a}.

On the one hand, a large set of software metrics have been proposed \cite{Hend96a}, the most known ones being the object-oriented metrics by Chidamber and Kemerer (CK) \cite{Chid94a}.
Although the CK metrics are widely used and valuable, they do not address the particularities of interfaces.
This was stated by Romano and Pinzger in their empirical studies for predicting change-prone interfaces \cite{Dani11a}.

On the other hand, Fowler and Beck \cite{Fowl99a} propose a set of bad smells in OO class design: \eg \emph{data class, god class, feature envy, duplicate code}.
They also propose refactorings for improving code quality with respect to the type of code smell.
Based on Fowler and Beck's definitions of class smells, several approaches to automatically improving code quality have been developed.
Murno proposes an approach based on code smells to automatically identify design smells and where to apply refactorings in a Software application \cite{Matthew2005}.
Trifu and Marinescu establish a clear distinction between OO structural problems and code smells, and present a causal approach to restructuring OO applications \cite{Trif05b}.
Liu \etal \cite{Liu12b} provide a deep analysis of the relationships among different kinds of bad smells and their influence on resolution sequences. 

Unfortunately, none of those code smells and OO metrics are applicable to Software interfaces, since interfaces do not contain any logic, such as method implementations, invocations, or attributes. Few recent works attempt to address the particularities of interfaces as shown in the next section.


\subsection{Interface Design Quality Metrics}\label{sec:relatedWorksInterface}

Boxall and Araban define a set of primitive counter metrics to measure the complexity and usage of interfaces \cite{Boxa04b}. 
Their metrics return the number of interface methods, all arguments of the interface methods, the interface client classes, \etc 
The authors in \cite{Abdeen2012} define more complex metrics that assess the interface design quality with regard to existing similarities among software interfaces, and with regard to the redundancy in interfaces hierarchies.

Martin proposed the Interface Segregation Principle (ISP) \cite{Mart00b} i.e., {\it ``do not design fat interfaces''}. Hence, an interface should group methods that are used together to serve a specific client. Large interfaces should be split into smaller  more specific ones so that clients will only have to know relevant methods without unwanted coupling to those that they do not use \cite{Mart00b,Dani11a}.
Ideally, an interface should not expose any unused methods and all the declared methods should be used by every client of the interface. 
With regard to the ISP principle, Romano and Pinzger \cite{Dani11a} used the {\it Service Interface Usage Cohesion} (SIUC) to measure the violation of ISP. 
SIUC is defined by Perepletchikov \cite{Pere07b}, and referred to by Romano and Pinzger \cite{Dani11a} as {\it Interface Usage Cohesion} (IUC):
\begin{equation}\label{equ:IUC}
\small
\centering
	\begin{aligned}
		IUC(i,c) & = \dfrac{num\_Used\_Methods(i,c)}{size(i)} &
		\\
		IUC(i)   & = \dfrac{\sum_c IUC(i,c)}{\mid clients(i) \mid} & \forall c\in clients(i)
	\end{aligned}
\end{equation} 
$IUC(i,c)$ computes the cohesion of the interface $i$ usage by a client class $c$, where $num\_Used\_Methods(i,c)$ returns the number of methods that are defined in $i$ and used by $c$, and $size(i)$ returns the number of all methods defined in $i$. 
$IUC(i)$ computes the usage cohesion of $i$ with regard to all its client classes. 
IUC states that an interface has a strong cohesion if every client class of that interface uses all the methods declared in it. 
IUC takes its value in the interval [0..1]. The larger the value of IUC, the better is the interface usage cohesion.  
Romano and Pinzger conclude that in order to limit changes propagation and facilitate software maintenance, the ISP should be respected when designing interfaces: \ie interfaces should be characterized by high values of IUC.

However, despite the success of this cited body of research effort on interface design metrics, to the authors' knowledge, there are no tools that help in identifying potential interface design anomalies and quantifying their impacts on interface design quality.

\subsection{Interface Design Quality Assessment}\label{sec:relatedWorksInterfaceQuality}
The authors in \cite{Dani11a} investigated the suitability of existing source code metrics to classify Software interfaces into change-prone and not change-prone. The metrics used in the study were Chidamber and Kemerer (CK) metrics, interface complexity and usage metrics and the IUC metric. They empirically evaluated their model for predicting change-prone Software interfaces by investigating the correlation between metrics and the number of changes in interfaces of ten Software open-source systems. 
The paper concluded that most of the CK metrics are not sound for interfaces and only perform well for predicting change-prone concrete and abstract classes. Therefore this confirms the claim that interfaces need to be treated separately. The IUC metric exhibits the strongest correlation with the number of interface changed. Hence IUC can improve the performance of prediction models for classifying Software interfaces into change-prone and not change-prone.

The work in \cite{cataldo2010impact} examined the relative impact of interface complexity (e.g. interface size and operation argument complexity) on the failure proneness of the implementation using data from two large-scale systems. This work provides empirical evidence that the increased complexity of interfaces is associated with the increased failure proneness of the implementation (e.g., likelihood of source code files being associated with a defect) and higher maintenance time. Our works goes much further by studying common interface design anomalies and investigating their impact on software quality.


\section{Background and Vocabularies}\label{sec:background}
As it is underlined in the introduction of this paper, Interfaces, unlike classes and abstract classes, are reference types used {\it to encode similarities which the classes of various types share}, but do not necessarily constitute a class relationship. 
Due to the type nature of interfaces, some interfaces could have/share similar semantics between  themselves (\ie define a shared subset of method declarations). 
For example, the interfaces {\it Set} and {\it Map} in the Java Collection Framework\footnote{Java tutorial, the core collection interfaces.}. 
{\it Set} declares in total 13 methods and {\it Map} declares 12 ones, where both declare the following shared subset of methods: {\it ``int size()''}; {\it ``boolean isEmpty()''}; and {\it ``void clear()''}. 
In fact, regardless of the differences between the implementations of those methods at classes that implement {\it Set} or {\it Map}, at interface level those methods are meant to present the same services: 
\begin{itemize}
\item {\it int size()} returns the size, as integer value, of the X collection. 
\item {\it boolean isEmpty()} returns whether the X collection is empty ({\it true} value) or not. 
\item {\it void clear()} removes all the items from the X collection. 
\end{itemize} 
Hence we say that {\it Set} and {\it Map} interfaces are, to certain extent, similar from the perspective of services/contracts they define. 
However, we argument that such a similarity between interfaces should be as low as possible, since interfaces are meant to encode shared services between classes, which in their turn define the implementation details of those services (\ie methods).

Interfaces are also meant \textit{to define the Application Programming Interfaces} (APIs) of frameworks and libraries (\eg Hibernate, jFreeChart and JHotDraw) which are meant to be extended or used by external client applications. 
As a result, those interface methods which define the APIs are not necessarily used by the framework/library itself. 
However, we argument that the presence of such interface methods, that are not used at all within the application itself, should be very limited, if any, within software tools that are not frameworks nor libraries (\eg Vuze, Opentaps, ArgoUML and Plantuml) --see \secref{sec:caseStudies} for more information about case-study applications. In fact, the presence of such interface methods within software tools means that the software interfaces are overused to declare unnecessary services, and/or the software interfaces were not well maintained during the software evolution.  

Before detailing the design anomalies, we introduce the vocabularies and notations we use in the rest of this paper.

	\myparagraph{Interface:} 
	we consider an interface as a set of method declarations (i.e., a set of signatures).   
	In this paper we consider only ``interfaces'' that declare at least one public method. 
	We do not take into account ``abstract classes'', ``marker interfaces'' or interfaces declaring only constants. 
	We consider only interface signatures (\ie methods) that are explicitly declared as ``public''. 
	We define the size of an interface $\interface$, $\iSize$, by the number of public methods that $\interface$ declares. 
	
	\myparagraph{Duplicate Interface Methods:}   
	we say that a signature $s$, that is declared within an interface $i$, is duplicated/cloned within another interface $i_d$ if $i_d$ declares another signature $s_d$ that is identical to $s$: \ie $s$ and $s_d$ have the same \textit{return-type}, the same \textit{name} and the same \textit{list-of-parameter-types}. 
	In such a case, we say that there is a {\it ``method clone''} between $i$ and $i_x$. We also refer to both methods $s$ and $s_d$ as {\it ``duplicate interface methods''}.\\    
	Hence we say that an interface $i$ is duplicated by another interface $i_x$ if all the $i$' methods are duplicated in $i_x$. 

	\myparagraph{Interface Implementations:} 
	the interface implementation classes are all the classes that belong to the interface sub-class hierarchy. 
	For example, if a class $a$ implements directly an interface $i_1$ which inherits from another interface $i_2$ ($i_1$ {\it `extends'} $i_2$), then $a$, and all the sub-classes of $a$, are implementation classes of both interfaces $i_1$ and $i_2$. 
	Each interface method is usually implemented (\ie overridden) by at least one public and concrete method defined in the interface implementation classes. 
	
	\myparagraph{Interface Client:} 
	An interface client is any class that calls any of the concrete methods overriding the interface methods and/or calls directly any of the interface methods --using the polymorphism mechanism.  
	We use $clients(\interface)$ to denote the set of $\interface$'s clients. 
	
	\myparagraph{Unused Interface Methods:} are public methods that are declared in an interface and are not called (none of themselves and/or of their implementations is called) by any client class. 
	Since the methods of interest are interface methods, we handle polymorphic method calls as follows:  
	a method $m_{called}$, that is declared in the interface $i_{receiver}$, is said called if within a class $c_{caller}$ there is an object $obj$ that has $i_{receiver}$, or any of its sub-classes or sub-interfaces, as declared type, and the message $m_{called}$ is sent to $obj$.    

\section{Interface Design Anomalies}\label{sec:designAnomalies}
In this section we characterise two common types of interface design anomalies: 
1) Interface methods duplication; 
2) Interfaces with unused methods.    
Then we illustrate the identified anomalies via real examples taken from well-known Software applications.

\subsection{Interface Methods Duplication}\label{sec:anomalyInterfaceSimilarity}
 Code clones are one of the most known bad smells in source code \cite{Fowl99a,Bali06a}. 
	Although interfaces do not provide implementations, code clones may still occur by duplicating signatures in several interfaces.   
	Such interfaces that share redundant signature declarations are thus similar from the point of view of public services/APIs they specify. 
	Hence, they indicate a bad organization of application services.

\begin{figure}[!h]
		\begin{center}
			\includegraphics[width=0.7\linewidth]{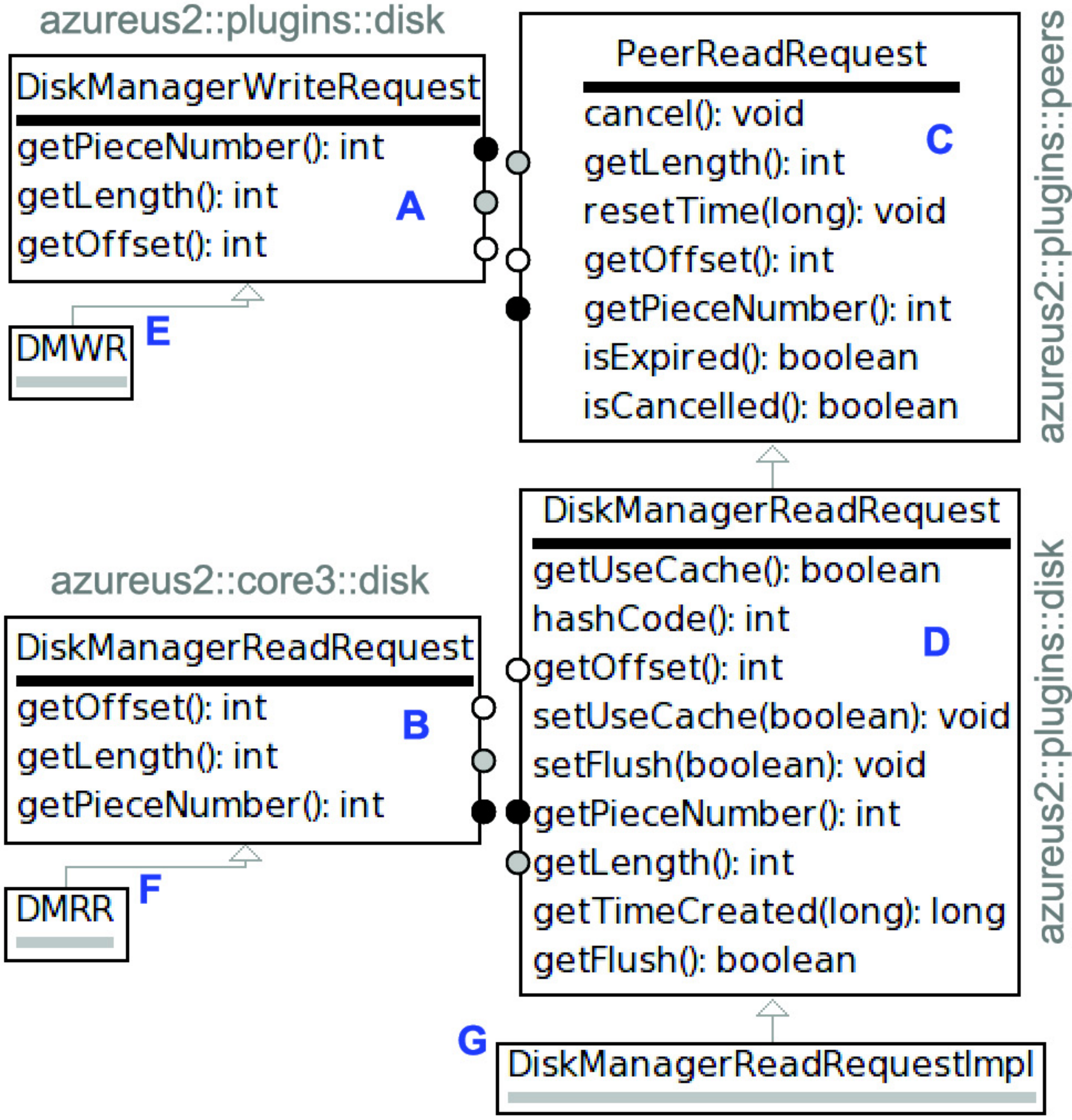}
			\vspace{-0.1cm}
			\caption{Example of Interface Similarity.
				\label{fig:interfaceSimilarity}}
		\end{center}
		\vspace{-0.2cm}
		{\footnotesize   
			Duplication of read \& write request methods in four interfaces in Vuze. 
		}
	\end{figure}

	\Figref{fig:interfaceSimilarity} shows an example of interface method duplication, taken from Vuze. 
    It shows a group of three signatures related to read and write request services that are duplicated in four interfaces. 
	Moreover, the figure shows that the interfaces (\A, \textit{DiskManagerWriteRequest}, and \B, \textit{DiskManagerReadRequest}) are identical since they declare the exact same set of those duplicated signatures.

	\myparagraph{Summary:}
		a conclusion is that the read and write request methods are declared in four interfaces of Vuze. 
		These duplicate declarations add more complexity and needlessly increase the size to the software system. 
		For example, to locate the read and write request methods in Vuze, one needs to locate four interfaces and inspect the sub-hierarchy of each of those interfaces. 
		Instead, those methods should be declared only once. 
		In the following, we propose a possible refectoring for reducing the declaration redundancy of read and write methods in Vuze. 
		First of all, we can safely remove interface \A (\textit{DiskManagerWriteRequest}) after moving all its incoming dependencies to interface \B (\textit{DiskManagerReadRequest}), \eg make the class \textit{DMWR} (\E) implement \B instead of \A. It is worth noting that classes \textit{DMWR} (\E) and \textit{DMRR} (\F) both provide identical implementations for those duplicate methods. 
		We then propose to rename \B to \textit{DiskManagerRWRequest} --since it declares what is apparently R\&W methods. 
		Then, we propose the following refactoring: 
		1) rename interface \D (\textit{DiskManagerReadRequest}) to \textit{...PeerReadRequest}; 
		2) make \C (\textit{PeerReadRequest}) a sub-interface of \B (\textit{DiskManagerReadRequest});
		3) remove the set of those duplicate signatures from both \C (\textit{PeerReadRequest}) and \D (\textit{DiskManagerReadRequest}). 
		The result of our proposed refactoring is that the R\&W methods are declared only in interface \B (\textit{DiskManagerReadRequest}). 
		Hence the number of interfaces is reduced --\A is removed; \E and \F, which need to implement the same set of methods, both implement \B; and the size of interfaces \C and \D is reduced.

\subsection{Interfaces with Unused Methods}\label{sec:anomalyInterfaceHarmfulness}
    One of the bad practices in designing interfaces is to design \textit{fat} interfaces that violate the Interface Segregation Principle (ISP) \cite{Mart00b}. A fat interface specifies a lot of \textit{generally useful methods}. 
	R. C. Martin \cite{Mart00b} stated that fat interfaces can be broken up into several interfaces where each interface serves a different set of clients. 
	Romano and Pinzger \cite{Dani11a} used the IUC metric to measure the violation of ISP, and stated that interfaces that has low values of IUC can cause difficulties in understanding and maintaining software systems. 
	
	Now, considering the existence of fat interfaces and focusing on their usage, \textit{what if some methods, that are declared in interfaces, are not invoked by any client class?}
    Since these methods are declared in the interface, they must be implemented, even though they are not actually used. Thus incurring additional cost in development time and effort and increasing code complexity. 
    Furthermore, since those methods do not serve the interface clients they thus violate the ISP design principle.  
	
	\begin{figure}[!h]
		\begin{center}
			\includegraphics[width=0.8\linewidth]{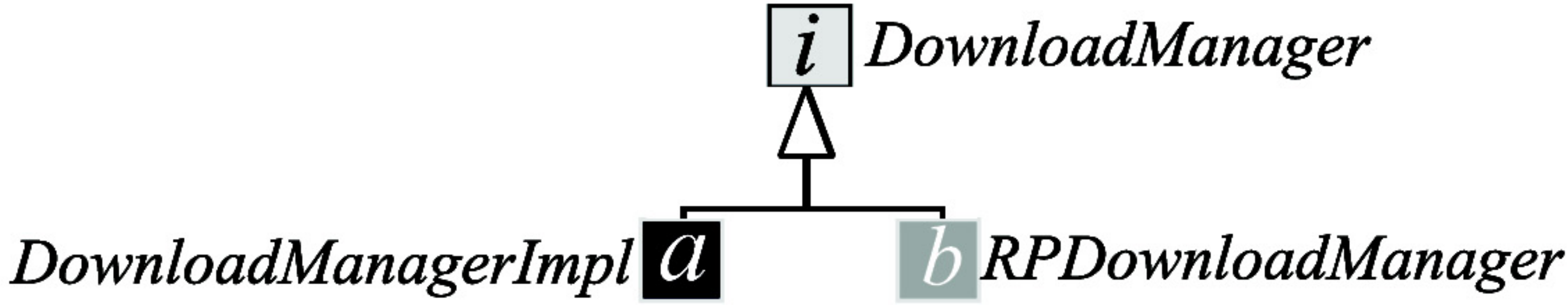}
            \caption{The sub-hierarchy of interface \texttt{plugins\-::download\-::Download\-Manager} ($i$), from Vuze.
            	\label{fig:interfaceHarmfulness}}
		\end{center}
	\end{figure}	
	
	For example, by inspecting interfaces in $Vuze$ application, we found the interface $plugins::download::DownloadManager$ which declares 31 public methods. 
	This interface is implemented by 2 classes (\Figref{fig:interfaceHarmfulness}): $DownloadManagerImpl$ ($a$) and $RPDownloadManager$ ($b$), both in $pluginsimpl::local::download$ package. 
	Both classes implement the 31 public methods that are declared in $DownloadManager$, with some additional public methods. 
	By inspecting the code, we found that 8 public methods of $DownloadManager$ are never invoked outside the implementing classes $DownloadManagerImpl$ and $RPDownloadManager$. 
	Those methods are: 
	\fvset{gobble=2}
	\begin{Verbatim}[fontshape=tt, frame=lines, fontsize=\small, label=Unused Methods of \textit{DownloadManger}]
		1) canResumeDownloads()
		2) canPauseDownloads()
		3) setSaveLocationManager(SaveLocationManager)
		4) removeDownloadWillBeAddedListener(...)
		5) addDownload(URL,URL)
		6) getGlobalDownloadEventNotifier()
		7) getSaveLocationManager()
		8) getDefaultSaveLocationManager()
	\end{Verbatim}
	Worst, we found that only the method {\tt getDefaultSaveLocationManager()} (8) is invoked inside the implementing classes: it is invoked inside $DownloadManagerImpl$. 
	As a consequence, there are 7 methods declared in $DownloadManager$, implemented twice (\ie 14 implementations), but never used. 
	By inspecting the implementation code of those 7 methods in $DownloadManagerImpl$ class, we found their total number of lines of code is 47 lines.
	
	\myparagraph{Summary:}
	as a conclusion, on one hand, the sub-hierarchy of interface $DownloadManager$ contains 14 unused public methods (47 lines of code in class $DownloadManagerImpl$ only) that can be safely removed. 
	On another hand, those methods can't be removed without removing their declarations in interface $DownloadManager$.

	Furthermore, {\tt getDefaultSaveLocationManager()} method (8) is declared at interface level as a public method, while it is used only inside implementing classes. 
	Good design dictates that it should be an internal private method rather than declared at interface level.    
	An interface must provide an abstract view of the public services provided its implementing classes. An interface should not leak the internal behavior of its implementing classes. 
	Otherwise, this will add ambiguity in system services and in the understanding of interactions among different subsystems \cite{Sark08x}. 
	

\section{Empirical Study Setup}\label{sec:setup}
This section describes the experiments conducted to study the interface design anomalies that we present in this paper through an exploratory study that aims at answering different research questions.

\subsection{Research Questions}\label{sec:questions}
\begin{Q}\label{q:AnomaliesPresence}
	To what extent duplicate / unused interface methods are present in real software applications?
\end{Q}
\begin{Q}\label{q:ProgramSize}
	To what extent duplicate / unused interface methods increase the size of software applications?
\end{Q}
\begin{Q}\label{q:InterfaceCohesion}
	To what extent duplicate / unused interface methods cause the degradation of the interface design quality particularly IUC?  
\end{Q}

\subsection{Assessment Measurements}\label{sec:measurements}
This section setups and describes the measurement we use later in \secref{sec:results} to answer the above research question. 

\vspace{+0.2cm}
\myparagraph{Presence of Duplicate and/or Unused Interface Methods.} 
For \refq{q:AnomaliesPresence}, we compute the ratio of the sum of all duplicate/unused interface methods (SDM/SUM) on the total number of interface methods within the concerned application. 
Let $DM(\interface)$ denote the number of methods that are declared in $\interface$ and duplicated (\ie re-declared) in other interfaces. 
Let $UM(\interface)$ denote the number of methods that are declared in $\interface$ and not used by any of the $\interface$' clients. 
Furthermore, let $NUM(\interface)$ denote the number of methods that are declared in $\interface$ and never used within the system, even not inside the interface implementation classes.  
The sum and ratio of duplicate/unused interface methods are defined as follows: 

\begin{equation}\label{equ:SDMandSUM}
\begin{aligned}
	SDM	& = \sum DM(\interface) 					& \forall \interface \in \interfaces
	\\
	SUM	& = \sum UM(\interface) 					& \forall \interface \in \interfaces
	\\
	SNUM& = \sum NUM(\interface) 					& \forall \interface \in \interfaces
	\\
	RSDM& = \dfrac{SDM}{\sum \iSize}				& \forall \interface \in \interfaces
	\\
	RSUM& = \dfrac{SUM}{\sum \iSize}				& \forall \interface \in \interfaces
	\\
	RSNUM&= \dfrac{SNUM}{\sum \iSize}				& \forall \interface \in \interfaces
\end{aligned}
\end{equation}

$RSDM$ and $RSUM$ represent respectively the ratio (percent) of duplicate and unused interface methods within the concerned software application. 

We expect that the values of $RSDM$ and $RSUM$ will be always very small and close to zero, particularly for $RSDM$. 
When the value of $RSDM$ increases and becomes notifiable this means that there is an undesired similarity between the application interfaces. 
As for $RSUM$, we expect that its values in software tools (\eg ArgoUML, Opentaps and Vuze), that are not frameworks nor libraries, should be close to zero. 
In the worst cases, the values of $RSUM$ for such software tools should be always smaller than for frameworks and libraries (\eg JHotDraw, Hibernate and jFreeChart). 
Moreover, if the values of $RSUM$ are close to those of $RSNUM$ this means that the software interface are overused for declaring unnecessary services which are useless and should be removed --except if those unused methods are APIs of a framework or library. 
Otherwise, if the values of $RSNUM$ are much smaller than those of $RSUM$ this means that the software interfaces are overused for describing the internal behavior of the implementation classes.

\vspace{+0.2cm}
\myparagraph{The Impact on Program Size.} 
For \refq{q:ProgramSize}, we estimate the Redundancy in interface methods by computing the difference between the sum of duplicate methods ($SDM$) and the virtual number of those methods if none of them was duplicated.  
Let $d\signatures$ returns the set of signatures of the duplicate interface methods where each signature is present only once in $d\signatures$.  
The size of $d\signatures$ would represent the number of duplicate method declarations if all method clones are eliminated from the interfaces. 
Thus, the difference between $SDM$ and the size of $d\signatures$ represents the Redundancy in Interface Methods ($ReIM$). 
Hence, we define the Ratio of Redundant Interface Methods ($RReIM$) as follows: 
\begin{equation}\label{equ:RReIM}
\begin{aligned}
	ReIM		& = SDM - \mid d\signatures \mid
	\\
	\\
	RReIM		& = \dfrac{ReIM}{\sum \iSize}				& \forall \interface \in \interfaces
\end{aligned}
\end{equation}

The value of $ReIM$ and/or $RReIM$ represents the additional size incurred to the software interfaces by the redundant declarations of interface methods. 
Ideally, the values of $ReIM$, as well of $RReIM$, are equal to zero, where none interface method is duplicate.  
The largest is the value of $ReIM$ and/or $RReIM$, the largest is the redundancy of the declarations of interface methods that are involved within duplication; thus the worst is the organization of services/APIs in software interfaces.   \\
As for unused interface methods, we compute the sum of lines of code of the concrete methods overriding the interface methods that are never used within the software system. 
We use $NULC$ (and $RNULC$) to denote the sum (and the ratio) of lines of code of those never used implementations.

\vspace{+0.2cm}
\myparagraph{The Impact on Interface Cohesion.}  
To assess the impact of duplicate/unused methods within a single interface $\interface$ on the cohesiveness of $\interface$ we need to evaluate the quality of $\interface$ with regard to the number of duplicate/unused methods that $\interface$ declares. 
We define the Ratio of Interface Duplicate/Unused Methods measurement (RDM/RUM) as follows: 

\begin{equation}\label{equ:RDMandRUM}
\begin{aligned}
	RDM(\interface) = \dfrac{DM(\interface)}{\iSize}
	\\
	\\
	RUM(\interface) = \dfrac{UM(\interface)}{\iSize}
\end{aligned}
\end{equation}

Where $DM(\interface)$ and $UM(\interface)$ returns respectively the number of duplicate and unused methods in $\interface$. 
Both measurements, $RDM$ and $RUM$, take their values in [0..1], where 1 is the worst value: \ie all the methods of $\interface$ are duplicated in other interfaces ($DM(\interface) = 1$) and/or unused within the software system ($UM(\interface) = 1$).\\
To characterize the impact of duplicate/unused interface methods on interface cohesion, we compute the correlation between the number (and ratio) of interface duplicate/unused methods ($DM$/$UM$ and $RDM$/$RUM$) and the interface usage cohesion ($IUC$), as defined in \equref{equ:IUC}. 
For this purpose, we make use of the Pearson’s correlation coefficient ($\rho$). 
This coefficient is generally used to express the strength of the relationship between two variables X and Y (in this paper X is $DM$/$UM$ or $RDM$/$RUM$, while Y is $IUC$). 
$\rho$ takes its values in [0..1]. 
If $IUC$ value increases when the number (ratio) of interface duplicate/unused methods increases, then we say that the number (ratio) of those methods is {\it positively} correlated with $IUC$. Thus, we say that duplicate/unused interface methods positively impact the interface cohesion (\ie they help in increasing the IUC value). 
However, if $IUC$ value decreases when the number (ratio) of interface duplicate/unused methods increases, then we say that duplicate/unused interface methods negatively impact the interface cohesion.

\subsection{Case Studies} \label{sec:caseStudies}
To study the interface design anomalies and answer the above questions, we conducted an empirical study on nine open-source Software projects:  
Six software tools and/or applications (1- GanttProject$_{v2.1}$, 2- Plantuml$_{v7935}$, 3- ArgoUML$_{v0.28.1}$, 4- Rapidminer$_{v5.0}$, 5- Opentaps$_{v1.5.0}$, 6- Vuze$_{v4700}$); 
two frameworks (7- JHotDraw$_{v7.1}$, 8- Hibernate$_{v4.1.4}$) and one library (9- jFreeChart$_{v1.0.14}$).  
We chose those applications since they contain a considerable number of interfaces and they differ in terms of functionality, number of interfaces, interface size, \etc (\tabref{table:basicInformation}). \\
We obtained those applications from sourceforge.net. 
We used the platform \textit{Moose} for data and software analysis \cite{Duca00b} to parse the application source-code and identify duplicate and unused interface methods. 
Note that the information we show is this paper about the studied applications are obtained after excluding the following: 
interfaces and classes related to the used programming language (\ie Java library);  
marker interfaces (\ie $|\iSize| = 0$); 
test-case classes (\ie classes inheriting from \textit{`JUnit TestCase'} class or packaged into \textit{`test(s)'} packages), and test interfaces (\ie interfaces that are packaged in \textit{`test(s)'} packages and/or all their implementations are test classes). 
Furthermore, to get precise information about unused interface methods, we consider only interfaces that have at least one implementing class (see the definition of ``Interface implementations'' in \secref{sec:background}).

\begin{table}[!h]
	\footnotesize
	\caption{Information about case-study software projects.}
	\vspace{-0.5cm}
	\label{table:basicInformation}
	\begin{center}
	\setlength\tabcolsep{2pt}
	\begin{tabular*}{1\linewidth}{@{\extracolsep{\fill}} ll@{\quad}rrr@{\quad}rrr}
	    \toprule
	    		&			&
	    				&			&			&
				\multicolumn{3}{c}{$\iSize$}\\
				\cmidrule(l{.05em}r{.5em}){6-8}	    
	    
	    		& Category	&
	    		$|C|$	&	LC		&	$|I|$	&
	    		min		&	max		&  sum \\
	    \midrule
	    GanttProject& 	{\it Proj. Mgment}		&
	    		886		&	44$k$	& 	106		&
	    		1 		& 	39		&  	587 \\
	    Plantuml&		{\it Dev. - Doc.}& 
	    		1224	&	76$k$	& 	122		&
	    		1 		& 	19		&  	358 \\
	    ArgoUML&		{\it Dev. - Design}		& 
	    		2202	&	166$k$	& 	126		&
	    		1 		& 	346		&  	1442 \\
	    Opentaps&		{\it Accounting}		& 
	    		3028	&	416$k$	& 	166		&
	    		1 		& 	153		&  	1598 \\
	    Rapidminer&		{\it Business}			& 
	    		3660	&	222$k$	& 	142		&
	    		1 		& 	48		&  	633 \\ 
	    Vuze&			{\it File sharing}		& 
	    		6564	&	635$k$	& 	933		&
	    		1 		& 	124		&  	6476 \\	    
	    \midrule		
	   	JHotDraw&		{\it Framework}			& 
	    		667 	&	65$k$	& 	46		&
	    		1 		&	49		&	537 \\
		Hibernate&		{\it Framework}			& 
	    		6195	&	373$k$	& 	471		&
	    		1 		& 	354		&  	3185 \\
	    jFreeChart&		{\it Library - Graphics}&
	   			590		&	160$k$	&	94		&
	   			1 		& 	134		&	560 \\
	    		
	    \bottomrule
	\end{tabular*}
	\end{center}
	{\footnotesize 
		\textsl{$|C|$ $\rightarrow$ number of classes (not interfaces); 
			$|I|$ $\rightarrow$ number of interfaces; 
			$sum$ $\iSize$ $\rightarrow$ number of all declared public methods in all interfaces.
		}
	}
\end{table}

\section{Results}\label{sec:results}
This section presents and analyses the results of the empirical study. 
Each subsection addresses one of the research questions that are outlined in \secref{sec:questions}.

\subsection{Presence of Duplicate and Unused Interface Methods}\label{sec:resultsUnusedAndDuplicatedMethods}
\begin{figure}[!h]
	\subfigure[%
			RSDM.%
		]{\label{sfig:ratiosOfDuplicateMethods}%
			\includegraphics[width=0.47\linewidth]{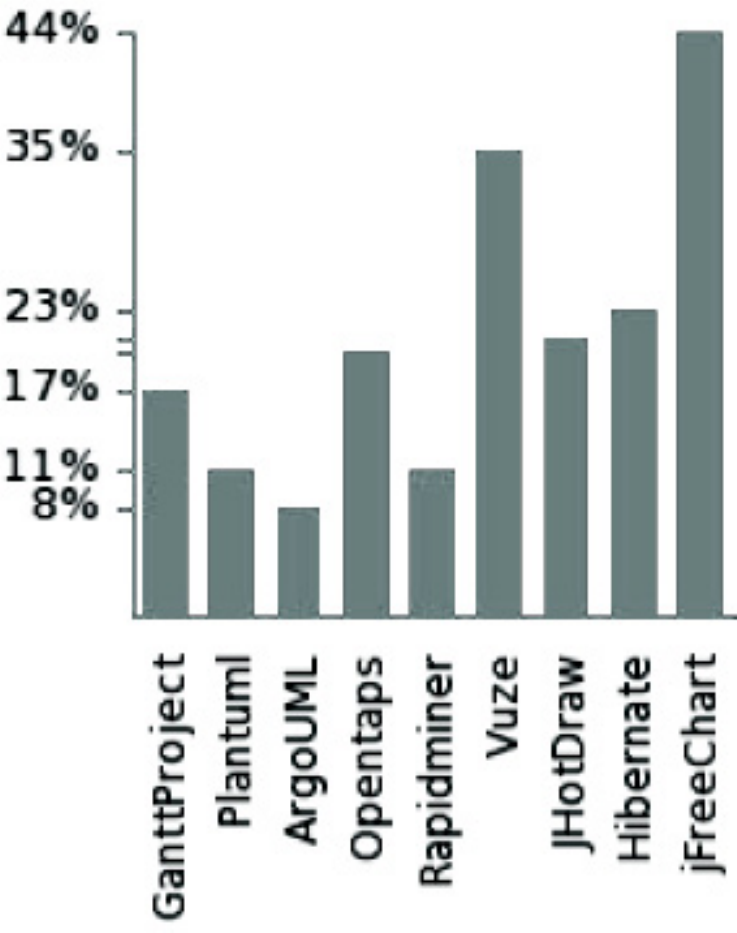}}
	\hfill
	\subfigure[%
			RSUM\&RSNUM ({\it blue} \& {\it cyan}).%
		]{\label{sfig:ratiosOfUnusedMethods}%
			\includegraphics[width=0.47\linewidth]{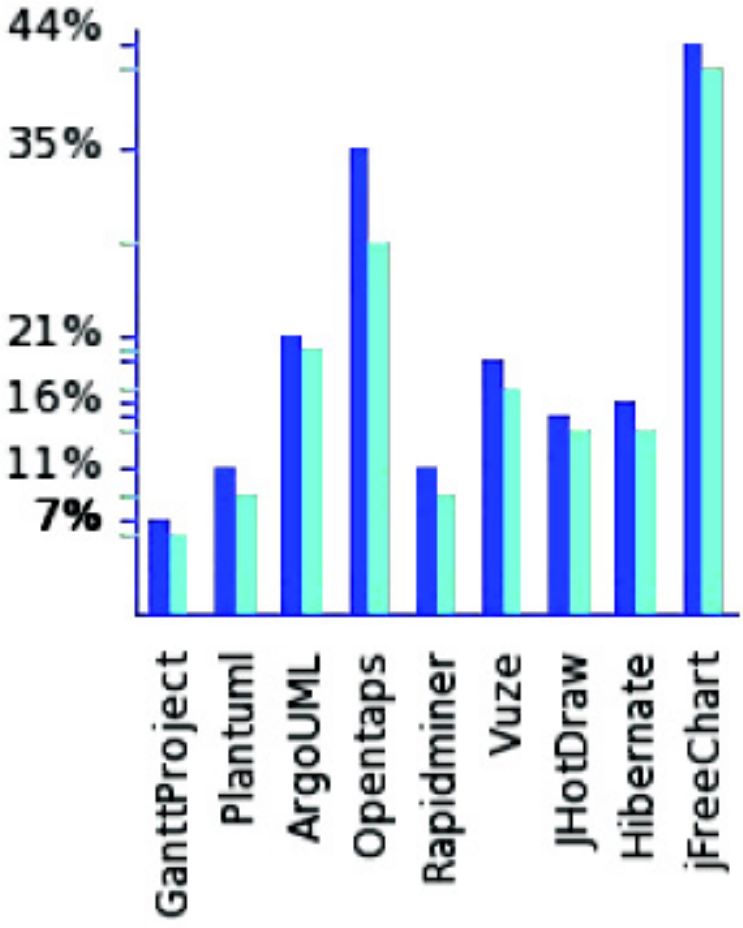}}
	\caption{Ratios, in percent, of duplicate and unused methods.\label{fig:duplicatedAndUnusedMethods}}
\end{figure}

To answer \refq{q:AnomaliesPresence},  \Figref{fig:duplicatedAndUnusedMethods} shows that unused and duplicate interface methods are  present, surprisingly to considerable extents, in all 9 studied applications. 
This statistically shows the uncomfortable evidence of the presence of the interface design defects that this paper presents. 
\Figref{sfig:ratiosOfDuplicateMethods} shows that 8\% (44\%) of interface methods are redundantly declared within the interfaces of ArgoUML (jFreeChart).
    
\Figref{sfig:ratiosOfUnusedMethods} shows that a considerable percent of the interface methods (7\% in GanttProject, about 20\% in ArgoUML and Vuze, and 35\% in Opentaps) are unfortunately unused outside the implementing classes; even though they are declared in interfaces and implemented as public methods.  
Furthermore \Figref{sfig:ratiosOfUnusedMethods} shows that almost all unused interface methods are never used. 
This can statistically mean that there is a very strong probability for an unused interface method to present a list of implementations (\ie concrete methods) that are not used at all and can be safely removed. 
This is in addition to the fact that methods to be used only inside their implementing classes should not be declared at the interface level.

\vspace{0.1cm}
\subsection{The Impact on Program Size}\label{sec:resultsUnusedMethodsComplexity}
To answer \refq{q:ProgramSize}, \Figref{sfig:ratiosOfUnusedMethods} shows that removing the declarations of unused methods from interfaces will considerably reduce interface size. 
Thus, this will considerably reduce the effort for understanding and recognizing the application services that are declared in the interfaces. 
Moreover, removing unused methods from interfaces helps in avoiding the blindly manner of implementing those methods in future classes. 

\begin{figure}[!b]
	\subfigure[%
			NULC.%
		]{\label{sfig:unusedLinesOfCode}%
			\includegraphics[width=0.47\linewidth]{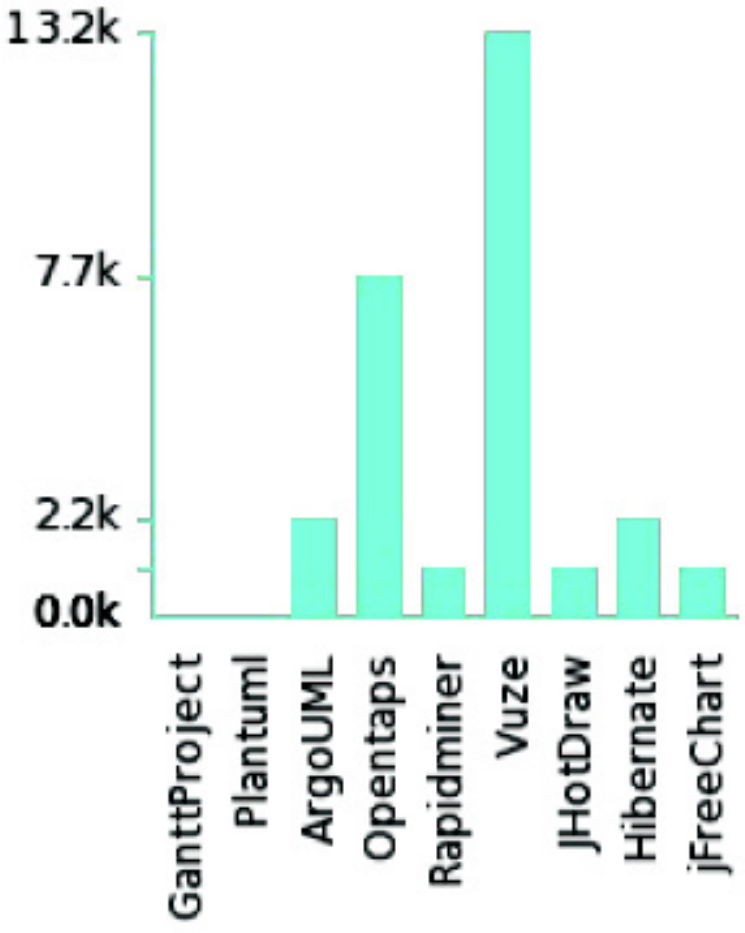}}
	\hfill
	\subfigure[%
			RNULC.%
		]{\label{sfig:ratioOfUnusedLinesOfCode}%
			\includegraphics[width=0.47\linewidth]{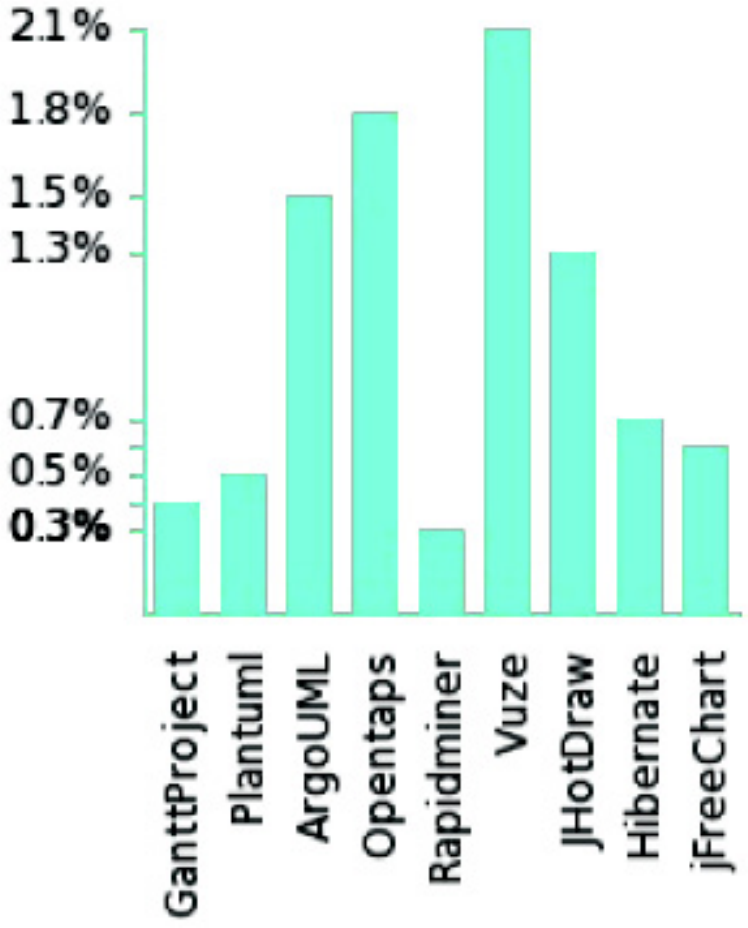}}
	\caption{The additional cost that is incurred by unused interface methods: number and ratio of never used lines of code.\label{fig:unusedImplementationsAndLinesOfCode}}
\end{figure}

Furthermore, \Figref{fig:unusedImplementationsAndLinesOfCode} shows that removing unused methods from interfaces will allow the software developer to remove safely a large number of unused implementations. 
Thus reducing the unneeded lines of code and consequently reduce the size and complexity of the implementation. 
For example removing about 7.7$k$ and 13.2$k$ unused LC respectively in Opentaps and Vuze, could be very meaningful. 
\Figref{sfig:ratioOfUnusedLinesOfCode} shows that removing those unused LC in ArgoUML, Opentaps and Vuze, will reduce more than 1\% of the whole size and complexity of the considered software system.

\begin{figure}[!t]
	\begin{center}
		\includegraphics[width=0.47\linewidth]{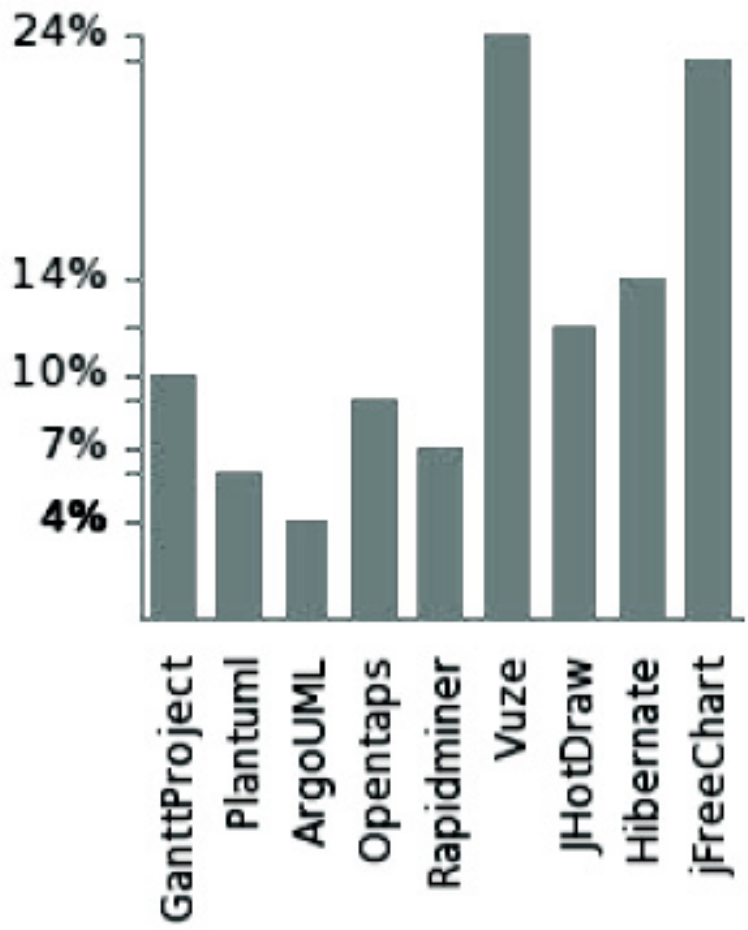}
		\vspace{-0.1cm}
		\caption{Ratio of redundancy in interface methods (RReIM): the extent to which duplicate methods are repiditly redeclared in different interfaces.\label{fig:RedundantMethods}}
	\end{center}
\end{figure}

To investigate the impact of method duplication on the size and comprehensibility of Software interfaces and the classes implementing them, \Figref{fig:RedundantMethods} shows the redundancy in method declarations within the software interfaces.

\Figref{fig:RedundantMethods} shows that to understand the services that are contracted between the application classes via interfaces, a considerable part of needed time will unfortunately be spent for rereading methods that are redundantly re-declared in the application interfaces. 
As example, for jFreeChart library, the maintainer might spend 20\% of his/her time for reading APIs  that he/she already read somewhere else in jFreeChart Interfaces. 
In ideal case where shared services between classes and/or the APIs are declared once, and only once, in the application interfaces, the interface size will then be considerable reduced, as demonstrated in \Figref{fig:RedundantMethods}.  
As a consequence, this will reduce the time and effort needed for understanding the application interfaces.

Furthermore, if the software developer is interested in one/some of those duplicate methods, he/she must capture all the interfaces declaring those methods and inspect their implementations separately to be able to decide the most appropriate method to use. 

\begin{figure}[!b]
	\begin{center}
		\includegraphics[width=0.45\linewidth]{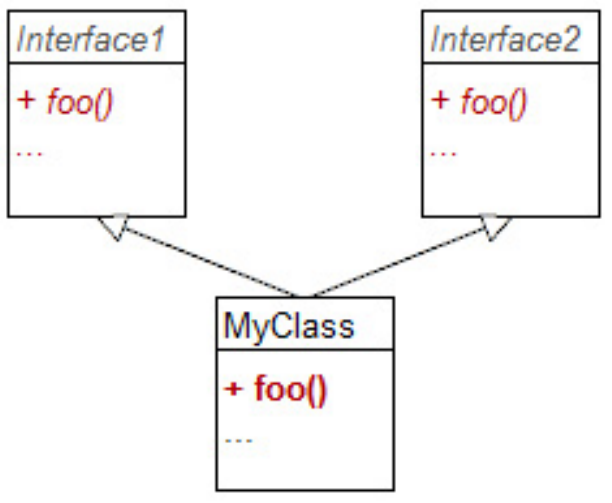}
		\vspace{-0.2cm}
		\caption{Method overriding from multiple interfaces ({\tt MyClass.foo()}).
		\label{fig:MethodOverrriding}}
	\end{center}
\end{figure}

Moreover, because of the multiple inheritance mechanism of interfaces, the maintainer might pass over some implementations several times since they represent implementations of different methods duplicate at the interface level (\eg \Figref{fig:MethodOverrriding}). 
Recall our example of interface method duplication that is taken from Vuze application \Figref{fig:interfaceSimilarity}. 
In this example, the three implementations (methods) of {\tt getOffset(), getPieceNumber()} and {\tt getLength()}, that are defined in {\it DiskManagerReadRequestImpl} class, are implementations of six different declarations coming from {\it DiskManagerReadRequest} and {\it PeerReadRequest} interfaces.      
However, this not the case of all the implementations of duplicate interface methods. 
For example, by inspecting the implementations of duplicate interface methods in Vuze, we found that only 1.3\% of them (59 ones) are shared between different duplicate interface methods: \ie each of those 59 implementations represents an implementation of several methods declared in Vuze interfaces. Of course the number/ratio of shared implementations between different duplicate interface methods differs among the case studies 
(\eg jFreeChart: 3.2\%; JHotDraw: 3.7\%; Hibernate: 5.2\%; GanttProject: 0.5\%; ArgoUML: 2\%; Rapidminer: 3.4\%).

\begin{figure}[!b]
	\subfigure[%
			$IUC(\interface)$ and $UM(\interface)$.%
		]{\label{sfig:correlationBetweenIUCAndUM}%
			\includegraphics[width=0.40\linewidth]{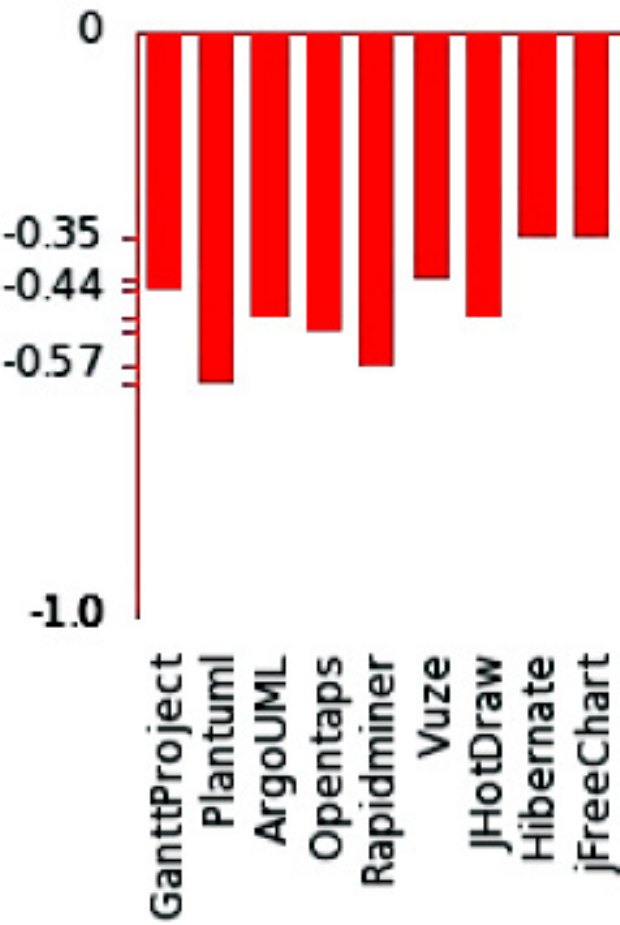}}
	\hfill
	\subfigure[%
			$IUC(\interface)$ and $RUM(\interface)$.%
		]{\label{sfig:correlationBetweenIUCAndRUM}%
			\includegraphics[width=0.40\linewidth]{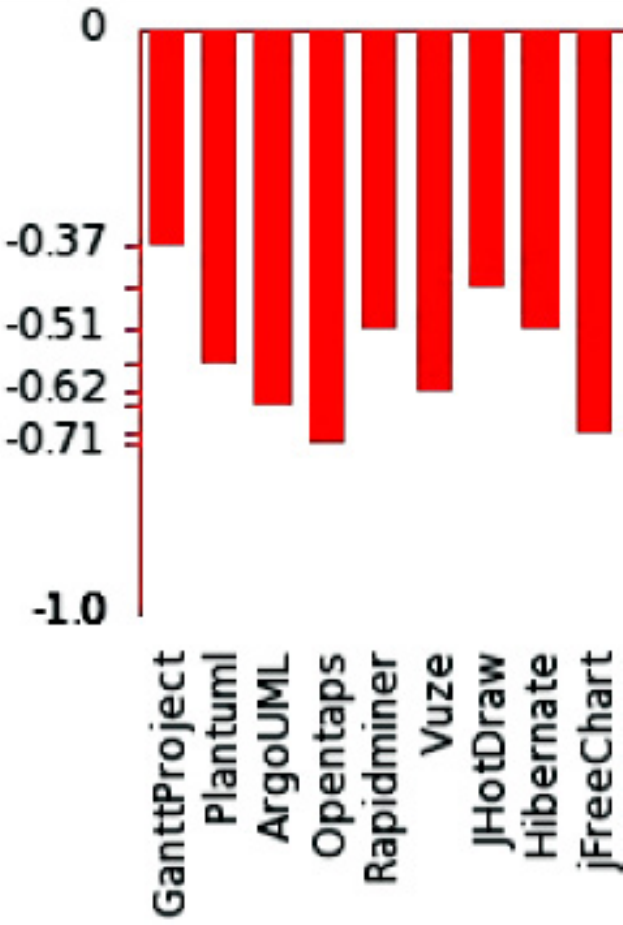}}
	\vspace{-0.2cm}
	\caption{The correlation between Interface Cohesion (IUC) and the number, and ratio value, of interface unused methods (UM and RUM).\label{fig:correlationIUCAndUM}}
\end{figure}

\begin{figure}[!b]
	\subfigure[%
			$IUC(\interface)$ and $DM(\interface)$.%
		]{\label{sfig:correlationBetweenIUCAndDM}%
			\includegraphics[width=0.38\linewidth]{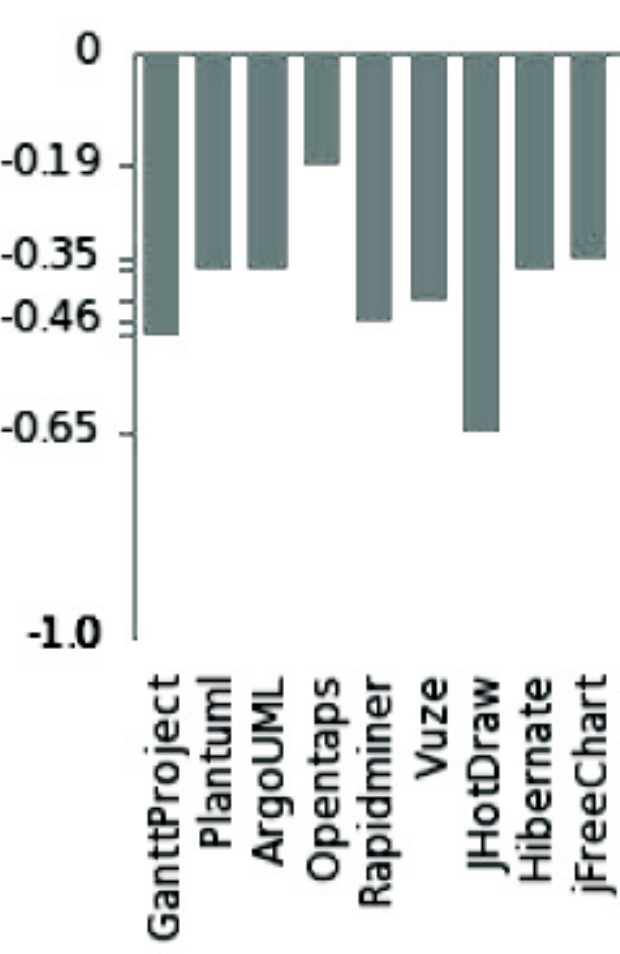}}
	\hfill
	\subfigure[%
			$IUC(\interface)$ and $RDM(\interface)$.%
		]{\label{sfig:correlationBetweenIUCAndRDM}%
			\includegraphics[width=0.38\linewidth]{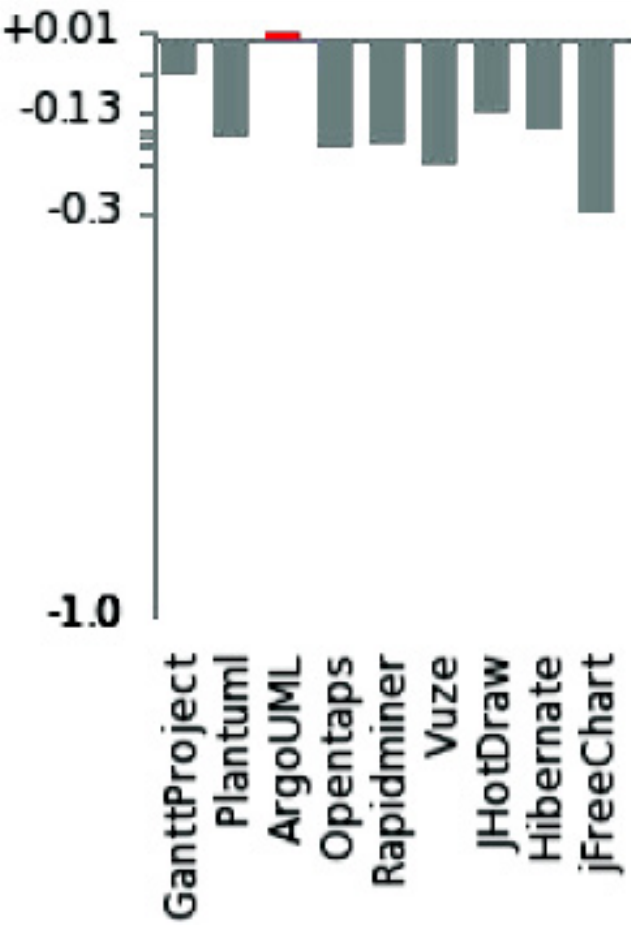}}
	\caption{The correlation between Interface Cohesion (IUC) and the number, and ratio value, of interface duplicate methods (DM and RDM).\label{fig:correlationIUCAndDM}}
\end{figure}

\subsection{The Impact on Interface Cohesion}\label{sec:resultsCohesionWithUnusedAndDuplicateMethods}
\Figref{fig:correlationIUCAndUM} shows that there is always a notifiable negative correlation between either the number or the ratio of unused interface methods and the interface usage cohesion. 
\Figref{sfig:correlationBetweenIUCAndUM} shows that for 7 case-studies, against 2 ones, there is a moderate negative correlation ($-0.4 > correlation \geq -0.6$) between the number of unused methods and the usage cohesion of an interface. 
This leads to the conclusion that unused methods, whether used or not inside the implementing classes, can have a negative effect on the interface cohesion. 
Those methods, even though they are not used by the interface clients, can drift the interface cohesion quality and the interpretation of IUC's value. 
This conclusion wins more evidence when comparing $IUC(\interface)$ values to $RUM(\interface)$ ones, since both measurements depend on $\interface$'s size and take their values in [0..1]. 
\Figref{sfig:correlationBetweenIUCAndRUM} shows that for 7 case-studies, against 2 ones, there is a moderate-to-strong negative correlation ($-0.5 > correlation \geq -0.71$) between interface cohesion and the relative number of unused methods that are declared into the interface. 

With this evidence, the $RUM$ measurement is important, not only for assessing the interface quality with regard to unused methods, but also to assist in the interpretation of $IUC$ values. 
In other terms, for interfaces that have low cohesion, it is advised to start by checking whether the $RUM$ values of those interfaces are high or not. 
If the highest the value of $RUM$, for a given interface $\interface$, the highest the probability that the $\interface$' cohesiveness is degraded because of the unused methods that are declared in $\interface$. 

As a summary to answer \refq{q:InterfaceCohesion}~: interfaces that are characterized by high values of $RUM$ are fort probably characterized by a small values of $IUC$. 
For sure, this finding does not claim that the interface cohesion is impacted only by the unused interface methods. 
Precisely, this finding outlines the negative impact of unused methods on the interface cohesiveness, whatever the size of the interface, small or large. 
Removing unused methods from those interface will improve their cohesion quality (\ie IUC value). 
Otherwise, for interfaces that declare APIs to be used by external client applications, this finding implies that is important to distinguish those APIs from other interface methods. 
This is important at to avoid erroneous conclusion about the interface cohesion.

Unlike unused methods, \Figref{fig:correlationIUCAndDM} shows that the correlation between duplicate methods and interface cohesion is not conclusive. 
It shows that the correlation between DM and IUC is classified between weak-to-moderate negative correlation, and that between RDM and IUC is poor for almost all case-studies. 
To answer \refq{q:InterfaceCohesion} we say that statistically the variation in DM account only for a small part of the variability in IUC. 
For precision, for 8 case-studies against only one, the correlation between IUC and DM is smaller than $-0.35$. 
Thus, the variation in DM account for {\it at least} 12\% ($-0.35^2$) of the variability in IUC. 
This means, statistically speaking, that method duplication in interfaces can cause, to limited extents, degradation of interface usage cohesion. 
Any way, what we could be sure about is that for all 9 case-studies there is always a {\it negative} correlation between IUC and the number of duplicate interface methods. 
As a consequence, duplicating method declarations in interfaces does not help to increase IUC --\ie duplicate methods, in the best cases, do not help to meet the ISP.

\section{Discussion}\label{sec:discussion}
We believe that the findings of the empirical study are interesting, particularly for software quality engineers and maintainers. 
Surprisingly, the study shows that the defect of duplicate methods in interfaces is widely present in well-known object-oriented applications.  
Thus, interfaces, if they are not well designed, can involve many redundant method declarations causing ambiguity in the organization of APIs. 
Because of this design defect and of multiple inheritance of interfaces, a concrete method implementation can be shared between several method declarations in different interfaces. 
This go against the role of interfaces --since it is the role of interfaces to represent shared services among classes. 
But unfortunately, duplicate methods in interfaces may lead to the fact that an implementation represent shared service among different interfaces. 
Furthermore, the study shows that the presence of duplicate methods in interfaces degrades the interface usage cohesion (IUC). 
As a consequence, and as demonstrated by Romano and Pinzger \cite{Dani11a}, this can violate the ISP principle \cite{Mart00b}, thus increase the effort needed to maintain those software systems that have interfaces with duplicate methods. 
This leads us to the following conclusion: 
{\it ``Do not duplicate methods in interfaces. Think twice before writing new interfaces and keep in mind that there are probably other interfaces that you can use and/or reuse.''}

Similar results were found for interfaces with unused methods.  
On one hand the study shows that unused methods degrades IUC. 
On another hand, surprisingly interfaces may declare many unused methods, introducing unnecessarily contracts between the software modules. 
Thus incurring unnecessarily complexity in the implementation and increased cost. 
Notice that among the studied applications there were only two frameworks (JHotDraw and Hibernate) and one library (jFreeChart). 
Thus one can argue that unused methods are APIs for expected plugins and build-on tools. 
What is unexpected, is that tools, such as ArgoUML, Opentaps and Vuze, all include interfaces with unused methods as in, or more than, those frameworks and library. 
This leads to the uncomfortable question: {\it ``Why interface designers declare unused methods, and what if those methods will not be used in future releases?''}. 
Interfaces are usually designed before classes and other implementations. 
Misunderstanding of costumer/system requirements and/or misplaced generosity of software designers can lead to declaring unnecessarily services in interfaces. 
Thus, if the designer is not fully sure that a service will be required between the software modules, she/he should not declare it in interfaces. 
One can argument that those unused methods are designed to be probably used for next releases of the software. 
We argument that when generously designing interfaces for future releases, it would be much better to distinguish the contracts for future releases from the current ones. 
Implementing all the services that one imagined for future releases means that we just reached the final release. 
{\it The designer should consider the number of lines of code, complexity and development time that will be saved by avoiding to implement unused interface methods}.

Furthermore, our findings regarding interfaces with duplicate and/or unused methods provide the evidence that it is important to use new metrics, such as the $RDM$ and $RUM$ metrics that we propose in this paper. 
This is to evaluate the quality of interfaces with regard to duplicate and unused methods. 
In addition, $RDM$ and $RUM$ can help maintainers in analyzing and interpreting $IUC$ values, where low values of $IUC$ can probably be caused by high values of $RDM$ and/or $RUM$ (\ie the concerned interface contains relatively large number of duplicate and/or unused methods).

\section{Threats to Validity}\label{sec:threats}
This section considers different threats to validity of our findings. 
We mainly discuss threats to {\it internal, external} and {\it conclusion} validity. 

First of all, the threats to internal validity of our study concern the independent variables that we used in our study, such as the values of the metrics (\eg $IUC$, $UM$ and $DM$). 
In our study, the values of all used independent variables are computed using static analysis tools and deterministic functions built on the Moose platform, that always return the same results. 
We well handled the problem of polymorphic methods calls, that is due to late-binding mechanism, as explained in \secref{sec:background} (see {\it ``Unused Interface Methods''} paragraph). 
Furthermore, our approach to detect method clones in interfaces is limited to compare interface signatures whether they are identical or not. 
Because of the absence of parameters' names within interface methods, the results that concern duplicate methods can be somewhat affected. 
For example, at class level, the methods {\tt \#add(String {\it name})} and {\tt \#add(String {\it webURL})} are not identical, even thought at interface level the signature of those methods is the same ({\tt \#add(String)}). 
Nevertheless, we believe that it is desirable to elaborate future studies on interface method clones that take into account the the parameters' names at implementation level and the semantic similarity between them.          

The threats to external validity concern the generalization of our findings. 
As a matter of fact, the external validity concerns usually arise from using a limited set of well-known software projects \cite{Wrig10a}. 
In our empirical study, this threat can be due to the fact that all case-studies are open-source software projects and they are developed in Java. 
Even though we carefully selected nine software systems that differ in size and utility, we believe that the generality of our findings should be verified for other industrial projects. 
Verifying our findings for projects that are developed in different programming languages, such as C++, is also desirable. 

As for threats to conclusion validity which concern the relationship between the treatment and the outcome, can be due to the use of statistical tests to support our findings. 
We used the Spearman correlation coefficient to investigate the relation between duplicate/unused interface methods and the interface cohesion (to answer \refq{q:InterfaceCohesion}).  
The Pearson’s correlation coefficient is sensitive only to a linear relationship between two variables. In cases when it is possible to have non-linear relationships, Pearson’s correlation will 
lead to an erroneous interpretation. We believe that this threat to validity is mainly related to our finding on the relationship between method clones in interfaces and interface cohesion --since the correlation was not conclusive.  

\section{Conclusion and Future Work}\label{sec:conclusion} 
In this paper, we identified and characterised key interface design anomalies and illustrated them via real examples taken from well-known open source applications. We concentrated on two common anomalies: interface methods duplication and interfaces with unused methods.
We conducted an empirical study covering 9 open source projects to evaluate the quality of interfaces using qualitative and quantitative analysis of the source code in order to quantify the presence of interface design anomalies and estimate their impact on the software quality attributes such as maintainability.
The results showed that the studied design anomalies are reliable symptoms of poor interface design. They are present, to different degrees, in interfaces of real systems.
Our findings suggest a strong need for researchers and engineers to distinguish between classes and interfaces when estimating the quality of software applications.
Additionally more interface-specific quality metrics are needed to measure violations of interface design principles particularly the Interface Segregation Principle (ISP). 
Moreover, software development tools should be enhanced to detect interface design anomalies then alert the developer and suggest appropriate refactoring. 
This can assist software engineers to avoid interface methods duplication so as to reduce the software complexity and wasted development time. 
In addition, tools can provide interface usage information to help reduce declaring unused methods or leaking the internal behavior of implementing classes. 
However further research is needed to achieve and validate this.
Our future work will concentrate on defining quality metrics to detect interface design anomalies and extract candidate interfaces for refactoring. 
Additionally we plan to investigate to which extent the existing software metrics can be used for detecting interface design anomalies. 
Further experiments on real software systems will be considered for validating the metrics. 
Another direction of future work is to investigate the associations between interface design defects and those of classes.
      

\section*{Acknowledgment}
This publication was made possible by NPRP grant \#09-1205-2-470 from the Qatar National Research Fund (a member of Qatar Foundation).

\bibliographystyle{IEEEtran}
\bibliography{IEEEabrv,localWithoutURLs}


\end{document}